\documentclass[11pt,a4paper]{article}
\usepackage{amsfonts,amsmath,amssymb}
\usepackage{graphicx}
\usepackage{latexsym}
\usepackage{hyperref}
\usepackage{setspace}
\usepackage{xcolor}
\usepackage{ragged2e}
\usepackage{cite}
\usepackage[labelfont=bf]{caption}
\usepackage{subcaption}
\usepackage[square,sort&compress,numbers]{natbib}
\usepackage[left=2.54cm, right=2.54cm, top=2.54cm, bottom=2.54cm]{geometry}
\date{}
\allowdisplaybreaks[3]
\usepackage{orcidlink}
\title{Domain Wall and Holographic Dark Energy in $f(Q)$ Theory of Gravity}
\author{S. P. Hatkar$^{1}$\orcidlink{0000-0002-3970-481X}, D. P. Tadas$^{2}$\orcidlink{0000-0002-1572-6213}, S. D. Katore$^{3}$\orcidlink{0000-0003-0521-4334}\\
	\small$^{1}$Department of Mathematics, A.E.S. Arts, Commerce and Science College, Hingoli-431513, India.\\
	\small$^{2}$Department of Mathematics, Toshniwal Arts, Commerce and Science College, Sengaon-431542, India.\\
	\small$^{3}$Department of Mathematics, Sant Gadge Baba Amravati University Amravati-444602, India.}
\date{}

\begin{document}
	\maketitle
	\onehalfspacing
\begin{abstract}
In the present research article, we have studied the flat FRW space time with domain wall and holographic dark energy in the context of $f(Q)$ theory of gravity. The exact solutions of the field equations are obtained using two different forms of varying deceleration parameters: (i) $q=-b(t)-1$ and (ii) $q=-lt+m-1$. It is observed that the universe is accelerating and expanding. Furthermore, we have also discussed some physical parameters of the investigated model.

\end{abstract}

\section{Introduction}
General Relativity (GR) is a very successful gravity theory, however it does not fully explain the universe's early and late-time acceleration. A modification to GR is required to explain the universe's accelerating expansion. Recent cosmological observations shows that the universe is expanding at an accelerated rate due to the dark energy (DE), an unknown form of energy with high negative pressure, which causes repulsion among galaxies. This expansion is confirmed by observations of type Ia supernovae \cite{Riess et al. 1998,Perlmutter et al. 1998,Perlmutter et al. 1999}, Cosmic Microwave Background (CMB) anisotropies, Wilkinson Microwave Anisotropy Probe (WMAP) \cite{Spergel et al. 2003}, and large-scale structures \cite{Bennett et al. 2003,Tegmark et al. 2004}. In order to explain DE and late-time acceleration, modifications of GR have received more and more attention in recent years. According to the WMAP, our universe is made up of around 73\% dark energy and 23\% dark matter, with only 4\% ordinary matter.

In recent years, the study of holographic dark energy (HDE) models has become an interesting field that has generated a lot of interest in understanding DE cosmological models. Recently, the holographic principle has been introduced into cosmology \cite{Li 2004} to track the DE content of the universe. This principle was first put forward by Hooft \cite{Hooft 1993} in the context of black hole physics. According to the holographic principle, the entropy of a system scales is not with its volume but with its surface area. Granda and Oliveros\cite{Granda and Oliveros 2008,Granda and Oliveros 2009} proposed a new infrared cut-off for HDE density, demonstrating that it represents accelerated universe expansion and is consistent with current observational data, as well as studying the correspondence between quintessence \cite{Barreiro et al. 2000} , tachyon \cite{Sen 2002}, k-essence \cite{Armendariz et al. 2001}, and phantom DE models \cite{Caldwell et al. 2003}. Saadat \cite{Saadat 2011} investigated the FRW cosmological model based on HDE and described DE density using the Chevallier-Polarski-Linder parametrization. The HDE model has been investigated by Sarkar \cite{Sarkar 2014} in the Bianchi type-I universe with a linearly varying deceleration parameter. Shaikh et al. \cite{Shaikh et al. 2023} investigated an oscillating locally rotationally symmetric (LRS) Bianchi-$II$ HDE cosmological model in the $f(R)$ gravity. Recently, Gupta et al. \cite{Gupta et al. 2023} investigate the Tsallis HDE and bulk viscous cosmological models using $f(Q)$ gravity and have obtained the solution using power-law expansion.

Topological defects such as domain walls, cosmic strings, and monopoles form due to spontaneous discrete symmetry breaking during cosmological phase transitions in the early universe. However, the study of these cosmic phase transition defects can be used to solve cosmological problems and explain the evolution of the universe \cite{Oliveira et al. 2005,Leite and Martins 2011}. Katore et al. \cite{Katore et al. 2015,Katore et al. 2016,Katore et al. 2021} explored several cosmological models with domain walls, including FRW, axially symmetric, Bianchi type-$II$, $VIII$, and $IX$, in the context of the $f(R,T)$ and $f(G)$ theories of gravity, indicating that domain walls may exist in the early universe and disappear with time. Tiwari et al. \cite{Tiwari et al. 2017} studied transit cosmological models in the $f(R,T)$ gravity theory and obtained solutions to field equations using a suitable scale factor that yielded time-dependent deceleration parameters. Saikawa \cite{Saikawa 2017} investigates a cosmological scenario in which unstable domain walls arise in the early universe and collapse in late time, which produces large amounts of gravitational waves. Bishi et al. \cite{Bishi et al. 2021} investigated the Friedmann-Lemaitre-Robertson-Walker (FLRW) universe in $f(R,T)=R+\alpha R^2+\lambda T$ gravity with domain walls and quark matter using two different Equation of State (EoS) forms with constant and variable deceleration parameters. Caglar et al. \cite{Caglar et al. 2023} have investigated Ruban space time with domain walls and strange quark matter in the $f(R,T)$ theory of gravity and GR. Recently, Maurya et al. \cite{Maurya et al. 2023} have used the arbitrary function $f(Q)=Q+\alpha \sqrt{Q}+2\Lambda$, where $\alpha$ and $\Lambda$ are the model-free parameter and the cosmological constant, respectively, to study the LRS Bianchi type-I string cosmological model in the f(Q) theory of gravity.

Now a days, several modifications to GR have been proposed to provide a natural gravitational alternative for DE. Among these modifications are $f(R)$ gravity \cite{Buchdahl 1970,Felice and Tsujikawa 2010,Nojiri and Odintsov 2011}, $f(T)$ gravity \cite{Ferraro and Fiorini 2007,Linder 2010}, $f(R,T)$ gravity \cite{Harko et al. 2011,Houndjo 2012}, $f(G)$ gravity \cite{Nojiri and Odintsov 2005,Bamba et al. 2010}, and many others. The study focuses on recent modifications based on a Lagrangian density, a general expression of the non-metricity scalar $Q$, known as the $f(Q)$ theory of gravity, which was recently proposed by \cite{Jimenez et al. 2018,Jimenez et al. 2020}, and has received a lot of attention in the recent year. Harko et al. \cite{Harko et al. 2018} have introduced a new theory in the context of metric-affine formalism that extends symmetric teleparallel gravity and coupled non-metricity $Q$ to the matter Lagrangian. They also explored cosmic applications, where the proposed approach provides an alternative to DE. Mandal et al. \cite{Mandal et al. 2020a,Mandal et al. 2020b} used cosmographic parameters to reconstruct the correct form of the $f(Q)$ function in the $f(Q)$ theory of gravity, determining the cosmic expansion history of the universe. They also studied energy conditions and restricted free parameters to current values consistent with the accelerated expansion. Khyllep et al. \cite{Khyllep et al. 2021} has explores Einstein's alternative $f(Q)$ theory, focusing on the power-law form of $f(Q)$ gravity, analyzing its behavior, revealing a deviation from the $\Lambda$CDM at perturbation levels. Narawade et al. \cite{Narawade et al. 2022,Narawade et al. 2023} studied accelerated cosmological models in $f(Q)$ gravity with non-metricity $Q$ and found that the EoS parameter for DE crosses the phantom split and the strong energy condition is violated at late times in both models using two different forms of $f(Q)$. Capozziello and D'Agostino \cite{Capozziello and D'Agostino 2022} explore non-metric gravity $f(Q)$ and its impact on dark energy effects, using rational Padé approximations to suggest $f(Q)=\alpha+\beta Q^n$ as the best approximation for accelerated expansion of universe. Agrawal et al. \cite{Agrawal et al. 2023} have shown the matter bounce scenario of the universe and explored the stability of the flat FLRW cosmological model in the context of symmetric teleparallel $f(Q)$ gravity.

\section{The $f(Q)$ theory of gravity}
The $f(Q)$ gravity theory, also known as symmetric teleparallel gravity, is an extension of GR in which $Q$ is a non-metricity scalar. Several studies in the literature have suggested that $f(Q)$ gravity is one of the most promising modified theories of gravity to explain cosmic observations. The action of non-metricity based $f(Q)$ gravity is given by \cite{Jimenez et al. 2020},
\begin{equation}\label{Eq:1}
	S=\int \frac{1}{2}f(Q) \sqrt{-g} d^{4}x + \int L_{m}\sqrt{-g} d^{4}x
\end{equation}
where $f(Q)$ is a non-metricity scalar function of $Q$, $g$ is the determinant of the metric tensor $g_{\mu\nu}$, and $L_m$ is the matter Lagrangian.\

As a result of the symmetry of $g_{\mu\nu}$, the contraction of the non-metricity tensor is given as
\begin{equation}\label{Eq:2}
	Q_{\alpha\mu\nu}\equiv \nabla_{\alpha} g_{\mu\nu}
\end{equation}
Here the non-metricity tensor gives rise to two independent traces, namely $Q_{\alpha}$ and $\tilde{Q}^{\alpha}$ as
\begin{equation}\label{Eq:3}
	Q_{\alpha}={{Q_{\alpha}}^{\mu}}_{\mu}, \tilde{Q}^{\alpha}={Q_{\mu}}^{\mu\alpha}
\end{equation}
Now, the non-metricity scalar $Q$ is given by \cite{Jimenez et al. 2018}
\begin{equation}\label{Eq:4}
	Q= - \frac{1}{4}Q_{\alpha\beta\mu}Q^{\alpha\beta\mu}+\frac{1}{2}Q_{\alpha\beta\mu}Q^{\beta\mu\alpha}+\frac{1}{4}Q_{\alpha}Q^{\alpha}-\frac{1}{2}Q_{\alpha}\tilde{Q}^{\alpha}
\end{equation}
The disformation tensor is define as
\begin{equation}
	L^{\alpha}_{\mu\nu}=\frac{1}{2}Q^{\alpha}_{\mu\nu}-{Q^{\alpha}_{(\mu \nu)}}
\end{equation}\label{Eq:5}
The field equation of $f(Q)$ gravity is obtained by varying the action \eqref{Eq:1} with respect to the metric tensor $g_{\mu\nu}$ as \cite{Anagnostopoulos et al. 2021}
\begin{equation}\label{Eq:6}
	\begin{split}
		\frac{2}{\sqrt{-g}}\nabla_{\alpha} \left\{\sqrt{-g}g_{\beta\nu}f_{Q} \left[-\frac{1}{2} L^{\alpha\mu\beta} -\frac{1}{8}\left(g^{\alpha\mu}Q^{\beta}+g^{\alpha\beta}Q^{\mu}\right) +\frac{1}{4}g^{\mu\beta}\left(Q^{\alpha}-\tilde{Q}^{\alpha}\right)\right] \right\} \\ + f_{Q} \left[-\frac{1}{2} L^{\mu\alpha\beta}-\frac{1}{8}\left(g^{\mu\alpha}Q^{\beta}+g^{\mu\beta}Q^{\alpha}\right)+\frac{1}{4}g^{\alpha\beta}\left(Q^{\mu}-\tilde{Q}^{\mu}\right)\right]Q_{\nu\alpha\beta}+\frac{1}{2}\delta^{\mu}_{\nu}f=T^{\mu}_{\nu},
	\end{split}
\end{equation}
where $T^{\mu}_{\nu}$ is the energy-momentum tensor and $f_{Q}\equiv\frac{\partial f}{\partial Q} $.

Also, the energy momentum tensor of Domain wall (DW) and Holographic dark energy (HDE) is given by
\begin{equation}\label{Eq:7}
	T_{\mu\nu}=(\rho_{d}+\rho_{h})(g_{\mu\nu}+u_{\mu}u_{\nu})+p_d u_{\mu}u_{\nu}
\end{equation}
with $u_{\mu}u^{\nu}=-1$, $\rho_{d}$, $p_d$ - represents the energy density, pressure of domain wall respectively and $\rho_{h}$ is holographic dark energy density.

The Friedmann-Robertson-Walker (FRW) models are the best for describing the current large-scale structure of the universe. The homogeneous and isotropic nature of the FRW model plays an important role in understanding the origin of the universe. We have considered the flat FRW spacetime as
\begin{equation}\label{Eq:8}
	ds^2= -dt^2+a(t)^2\left(dx^2+dy^2+dz^2\right),
\end{equation}
where $ a(t) $ is scale factor. Recently, Pradhan et al. \cite{Pradhan et al. 2021} have studied dark energy behaviour of the flat FLRW universe with viscous fluid in $f(Q)$ theory of gravity. To investigate the behaviour of different cosmological parameters in terms of redshift z, we first define the scale factor $a$ in terms of redshift $(z)$ as $a(z)=\frac{1}{1+z}$, and we get the Hubble parameter $H$ as
\begin{equation}\label{Eq:9}
	H=\frac{\dot{a}}{a}
\end{equation}
The two different forms of HDE, which are found in the literature. The first form of HDE considered by Granda and Oliveros \cite{Granda and Oliveros 2008} as
\begin{equation}\label{Eq:10}
	\rho_{h}=3M_{p}^{2}\left(\zeta H^{2}+\eta \dot{H}\right)
\end{equation}
where $ \zeta, \eta$ are constants and $ M_{p} $ is the reduced plank mass with $ M_{p}^{-2}=1 $.\\
Also, the another form of HDE is defined by Saadat \cite{Saadat 2011} as
\begin{equation}\label{Eq:11}
	\rho_{h}=3d^{2}H^{2}
\end{equation}
where $ d $ is constant.\
\section{Field equation and Solutions}
The field equation \eqref{Eq:6} of $f(Q)$ gravity are obtained by using equations \eqref{Eq:7} and \eqref{Eq:8} as 
\begin{equation}\label{Eq:12}
	-6H^2f_{Q}+\frac{1}{2}f=-p_d
\end{equation}
\begin{equation}\label{Eq:13}
	-2\dot{H}f_{Q}-2H\dot{f_{Q}}+\frac{1}{2}f=\rho_{d}+\rho_{h}
\end{equation}
where the dot represents derivative with respect to time $t$. The field equations \eqref{Eq:12} and \eqref{Eq:13} mentioned above are systems of two equations with $f$, $p_d$, $\rho_d$, and $\rho_h$ as the four unknowns. In order to find the exact solutions to the field equation, we need two more conditions.\\
In this paper, we first assume that the $f(Q)$ model, as proposed by Harko et al. \cite{Harko et al. 2018}, takes a specific form with a linear and non-linear non-metricity scalars $(Q)$ as
\begin{equation}\label{Eq:14}
	f(Q) = \alpha Q + \beta Q^{n},
\end{equation}
where $\alpha$, $\beta $, and $n\ne1$, are free parameters. Solanki et al. \cite{Solanki et al. 2022} explored the DE evolution in $f(Q)$ gravity from spacetime geometry and found that geometrical generalization of GR can provide a convincing explanation for DE origin. The $f(Q)$ theory of gravity has been proposed as an alternative theory to general relativity to address issues such as the need for dark energy to explain the accelerated expansion of the universe. The choice of deceleration parameter $q$ is motivated by reproducing observed acceleration or providing a viable alternative explanation within modified gravity. It quantifies the rate at which the expansion of the universe slows down due to the gravitational attraction of matter and other energy components.
The deceleration parameter $q$ is define as 
\begin{equation}\label{Eq:15}
	q=-\frac{a\ddot{a}}{\dot{a}^2}
\end{equation}
where $a$ is the scale factor.

Secondly, for the solution of field equations, we have considered the two different form of linearly varying deceleration parameters such as (i) $q=-b(t)-1$ and (ii) $q=-lt+m-1$.
\subsection{Case-I:}
As we know, the sign of the deceleration parameter determines whether the universe is decelerating or accelerating. We are interested in the cosmic expansion of the universe. Recently, Akarsu and Dereli \cite{Akarsu and Dereli 2012} have proposed linearly varying deceleration parameter  as follows:
\begin{equation}\label{Eq:16}
	q=-lt+m-1
\end{equation}
The given equation leads to the following three cases:
\begin{itemize}
	\item [i)] If $l=0, m=0 \Rightarrow q=-1;$
	\item [ii)] If $l=0, m>0 \Rightarrow q=m-1;$
	\item [iii)] If $l>0, m>0 \Rightarrow q=-lt+m-1.$
\end{itemize}
If $q>0$, it represents decelerating expansion; if $q=0$, the expansion occurs at a constant rate; if $-1<q<0$, it indicates the power law expansion; if $q=-1$, it shows exponential expansion; and if $q<-1$, super exponential expansion occurs. They have studied all three cases. This new generalized linearly varying deceleration parameter covers Berman’s law and suggests that the result is consistent with cosmological observations. Also, Mishra et al. \cite{Mishra et al. 2016} have considered the new form of variable deceleration parameter as $q=b(t)$ with the appropriate choice of scale factor $a(t)=(sinh(\alpha t))^1/n$, where $\alpha$ and $n$ are constants. However, the possibility of other forms of deceleration parameter may give more appropriate consistency with observational data. Keeping in the view of linearly varying deceleration parameter form proposed by Akarsu and Dereli \cite{Akarsu and Dereli 2012}, Mishra et al. \cite{Mishra et al. 2016}, we would like to consider the following form of deceleration parameter as
\begin{equation}\label{Eq:17}
	q=-b(t)-1
\end{equation}
where $b(t)$ is the function of $t$ only. This is more general form of deceleration parameter than that assumed by Akarsu and Dereli \cite{Akarsu and Dereli 2012}. When $b(t)= lt-m$, we get the linearly varying deceleration parameter assumed in \cite{Akarsu and Dereli 2012}.\\
Here we take $b(t)= -\sec h^2t$ which leads to scale factor $a$, Hubble parameter $H$ and deceleration parameter $q$ are as follows:
\begin{equation}\label{Eq:18}
	a(t)= \sin ht
\end{equation}
\begin{equation}\label{Eq:19}
	H(t)=\cot ht 
\end{equation}
\begin{equation}\label{Eq:20}
	q(t)=\sec h^2t-1
\end{equation}
\begin{figure}[hbt]
	\centering
	\includegraphics[width=0.65\linewidth]{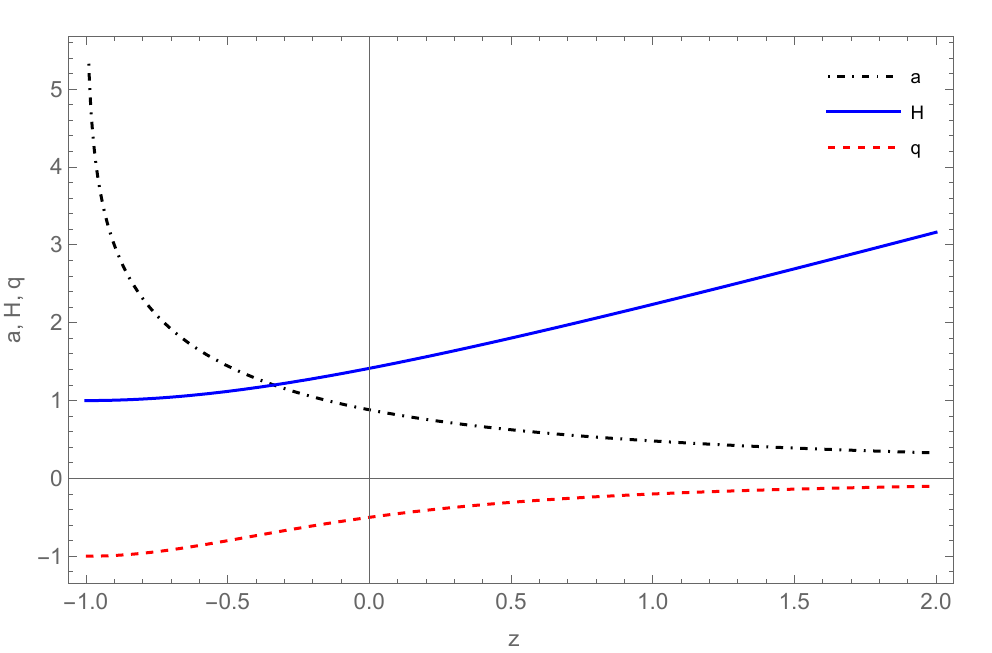}
	\caption{Plot of scale factor $(a)$, Hubble parameter $(H)$ and deceleration parameter $(q)$ \textit{vs.} redshift $z$}
	\label{fig:1}
\end{figure}
From figure \ref{fig:1}, it is observed that the deceleration parameter $q$ is negative throughout the evolution of the universe. At the present, $q=-0.5$, and at the latter time, it tends to $-1$, which is consistent with an accelerating universe and is supported by the observations of the Type Ia supernovae \cite{Riess et al. 1998,Perlmutter et al. 1999} and WMAP data \cite{Spergel et al. 2003}. The scale factor $a$ is increasing, which is as per the standard cosmological model of the universe. The Hubble parameter $H$ tends to constant as z tends to negative.\\
Solving equation \eqref{Eq:12} by using the values from equations \eqref{Eq:14} and \eqref{Eq:19}, the pressure $p_{d}$ of domain wall becomes
\begin{equation}\label{Eq:21}
	p_{d}=3 \alpha  \coth ^2(t) +\frac{1}{2} \beta  (2 n-1) 6^n \coth ^{2 n}(t)
\end{equation}
From equation \eqref{Eq:19}, the HDE $\rho_{h}$ defined by Saadat \cite{Saadat 2011} is obtained as
\begin{equation}\label{Eq:22}
	\rho_{h}=3 d^2 \coth ^2(t)
\end{equation}
Also, Using equation \eqref{Eq:19}, the HDE $\rho_{h}$ considered by Granda and Oliveros \cite{Granda and Oliveros 2008} is calculated as
\begin{equation}\label{Eq:23}
	\rho_{h}=3\left[(\zeta-\eta)\coth ^2(t)+\eta\right]
\end{equation}
\begin{figure}[hbt]
	\centering
	\includegraphics[width=0.65\linewidth]{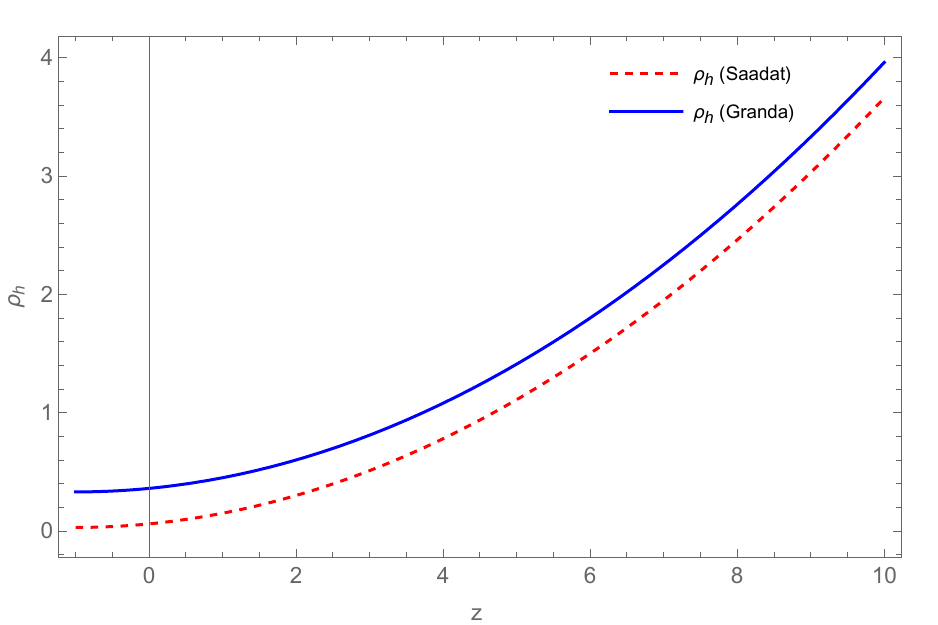}
	\caption{The plot of holographic dark energy density $\rho_h$ \textit{vs.} redshift $z$}
	\label{fig:2}
\end{figure}
The graphical representation of HDE density $\rho_{h}$ shows that it is initially high in past and gradually approaches to zero in present, as depicted in figure \ref{fig:2}. It has been found that the behaviour of both types of HDE is the same.\\
Using equation \eqref{Eq:22}, the DW energy density $\rho_{d}$ considered by Saadat \cite{Saadat 2011} is obtained as
\begin{equation}\label{Eq:24}
	\begin{split}
	\rho_{d}=&-2 \alpha +\left(5 \alpha -3 d^2\right) \coth ^2(t)+ \left(4 n^2-2 n+3\right)\beta  6^{n-1} \coth ^{2 n}(t)\\ &- \left(4 n^2-2 n\right)\beta 6^{n-1} \coth ^{2 n-2}(t)
	\end{split}
\end{equation}
Also, from equation \eqref{Eq:23}, the energy density of DW $\rho_{d}$ defined by Granda and Oliveros \cite{Granda and Oliveros 2008} is calculated as
\begin{equation}\label{Eq:25}
	\begin{split}
		\rho_{d}=&-(2 \alpha + 3 \eta )+\left[5 \alpha -3 (\zeta -\eta )\right]\coth ^2(t) +\left(4 n^2-2 n+3\right) \beta  6^{n-1}  \coth ^{2 n}(t) \\
		&-\left(4 n^2-2 n\right)\beta   6^{n-1} \coth ^{2 n-2}(t)
	\end{split}
\end{equation}
\begin{figure}[hbt]
	\centering
	\includegraphics[width=0.65\linewidth]{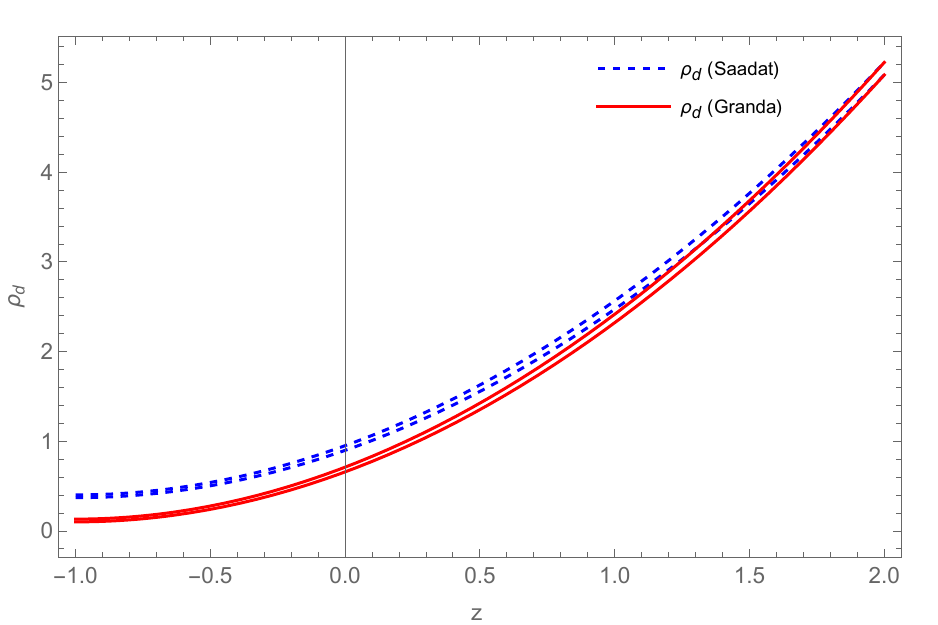}
	\caption{The plot of energy density of domain wall $\rho_d$ \textit{vs.} redshift $z$}
	\label{fig:3}
\end{figure}
From figure \ref{fig:3}, we observe that the energy density of the domain wall is positive and decreasing. It was large in the past and tends to zero as $z \rightarrow - 1$. The behaviour of Saadat type and Granda and Oliveros type HDE is alike.\\
Now, using equations \eqref{Eq:22} and \eqref{Eq:24}, the tension of the domain wall $\sigma_{d}$ regarding Saadat type HDE is obtained as
\begin{equation}\label{Eq:26}
	\begin{split}
		&\sigma_{d}= \left[\left(5-\frac{8}{\gamma }\right) \alpha -3 \left(1-\frac{1}{\gamma }\right) d^2 \right] \coth ^2(t) -2 \alpha \left(1-\frac{1}{\gamma }\right) \\
		&+\left[(4 n^2-2 n+3)-\frac{4 n^2+4 n}{\gamma }\right] \beta 6^{n-1} \coth ^{2 n}(t) - \left(1-\frac{1}{\gamma }\right) \left(4 n^2-2 n\right) \beta 6^{n-1} \coth ^{2 n-2}(t)
	\end{split}
\end{equation}
Also, from equations \eqref{Eq:23} and \eqref{Eq:25}, the domain wall tension $\sigma_{d}$ for Granda and Oliveros type HDE is calculated as
\begin{equation}\label{Eq:27}
	\begin{split}
		& \sigma_{d}= \left[\left(5-\frac{8}{\gamma }\right) \alpha -3(\zeta -\eta ) \left(1-\frac{1}{\gamma }\right) \right] \coth ^2(t) -(2 \alpha +3 \eta ) \left(1-\frac{1}{\gamma }\right) \\
		&+\left[(4 n^2-2 n+3)-\frac{4 n^2+4 n}{\gamma }\right] \beta 6^{n-1} \coth ^{2 n}(t)-  \left(1-\frac{1}{\gamma }\right) \left(4 n^2-2 n\right) \beta 6^{n-1} \coth ^{2 n-2}(t)
	\end{split}
\end{equation}
\begin{figure}[hbt]
	\centering
	\includegraphics[width=0.65\linewidth]{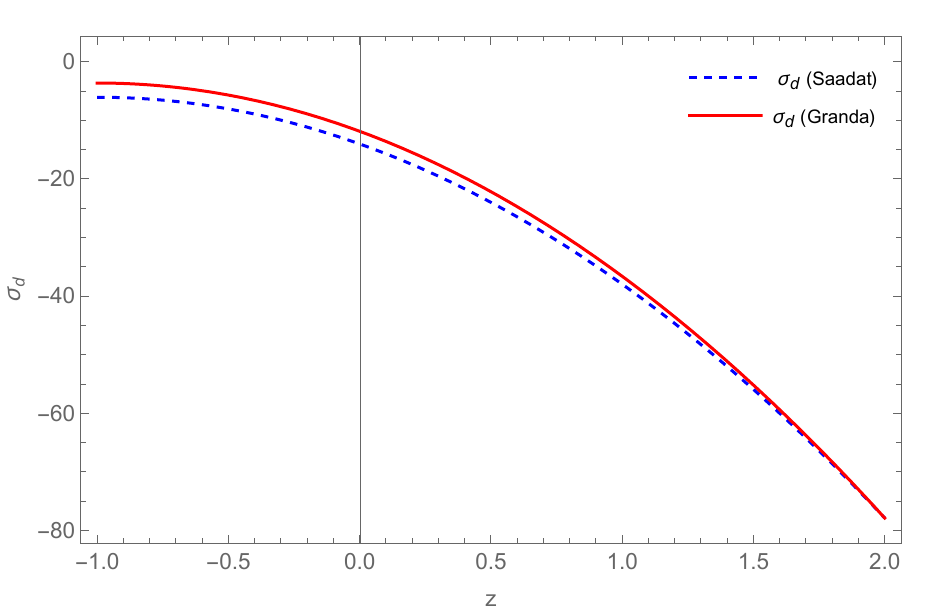}
	\caption{The plot of tension of domain walls $\sigma_d$ \textit{vs.} redshift $z$}
	\label{fig:4}
\end{figure}
The tension of the domain wall is negative throughout the evolution of the universe depicted in figure \ref{fig:4}. The negative tension of the domain wall means it behaves like invisible matter. The universe is dominated by HDE. It should be noted that, according to Zeldovich et al. \cite{Zel'Dovich et al. 1974}, the domain wall exists in the early epoch and vanish at the present. In our earlier works in $f(R,T)$ gravity, we found that for a flat universe, the domain wall could exist in the early universe and vanish in the latter stage of the evolution of the universe \cite{Katore et al. 2016}. Here the domain wall is an invisible matter; it may be an effect of HDE of $f(Q)$ gravity.

\subsection{Case-II:}
As discussed above, we have now solved the field equations by considering the linearly varying deceleration parameter proposed by Akarsu and Dereli \cite{Akarsu and Dereli 2012} in equation \eqref{Eq:16} with $l>0$ and $m>0$ as constants,
\begin{equation}\label{Eq:28}
	q=-lt+m-1.
\end{equation}
After solving above equation leads to
\begin{equation}\label{Eq:29}
	a(t)= c_{1} \exp\left[\frac{2}{\sqrt{m^2-2c_{2}l}} \arctan h\left(\frac{lt-m}{\sqrt{m^2-2c_{2}l}}\right)\right]
\end{equation}
where $c_{1}$, $c_{2}$ are constants of integration. Using equation \eqref{Eq:29}, the Hubble parameter $(H)$ becomes
\begin{equation}\label{Eq:30}
	H(t)=\frac{2}{-lt^2+2mt-2c_{2}}
\end{equation}
\begin{figure}[hbt]
	\centering
	\includegraphics[width=0.65\linewidth]{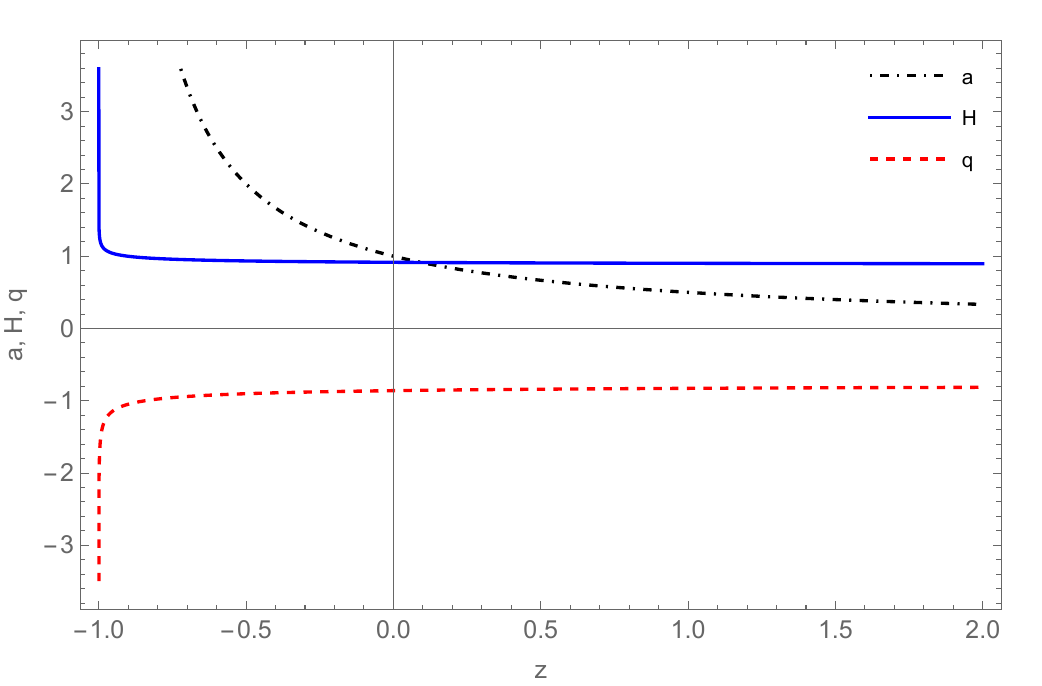}
	\caption{Plot of scale factor $(a)$, Hubble parameter $(H)$ and deceleration parameter $(q)$ \textit{vs.} redshift $z$}
	\label{fig:5}
\end{figure}
From figure \ref{fig:5}, it is clear that the deceleration parameter $q$ is negative throughout the evolution of the universe, i.e., the universe is accelerating. The Hubble parameter is constant and increases as $z \rightarrow -1$, i.e., the rate of expansion of the universe is increasing. The scale factor is also increasing.\\
Now, using the equations \eqref{Eq:14} and \eqref{Eq:30}, the equation \eqref{Eq:12} becomes
\begin{equation}\label{Eq:31}
	p_{d}=3\alpha \left(\frac{2}{-lt^2+2mt-2c_{2}}\right)^2 + \beta\left(\frac{2n-1}{2}\right) 6^{n} \left(\frac{2}{-lt^2+2mt-2c_{2}}\right)^{2 n}
\end{equation}
From equation \eqref{Eq:30}, the HDE $\rho_{h}$ for Saadat type \cite{Saadat 2011} is obtained as
\begin{equation}\label{Eq:32}
	\rho_{h}=\frac{12 d^2}{\left(-lt^2+2mt-2c_{2}\right)^2}
\end{equation}
Also, Using equation \eqref{Eq:30}, the HDE $\rho_{h}$ for Granda and Oliveros type \cite{Granda and Oliveros 2008} is calculated as
\begin{equation}\label{Eq:33}
	\rho_{h}=\frac{12\left[\zeta+\eta(lt-m)\right]}{\left(-lt^2+2mt-2c_{2}\right)^2}
\end{equation}
\begin{figure}[hbt]
	\centering
	\includegraphics[width=0.65\linewidth]{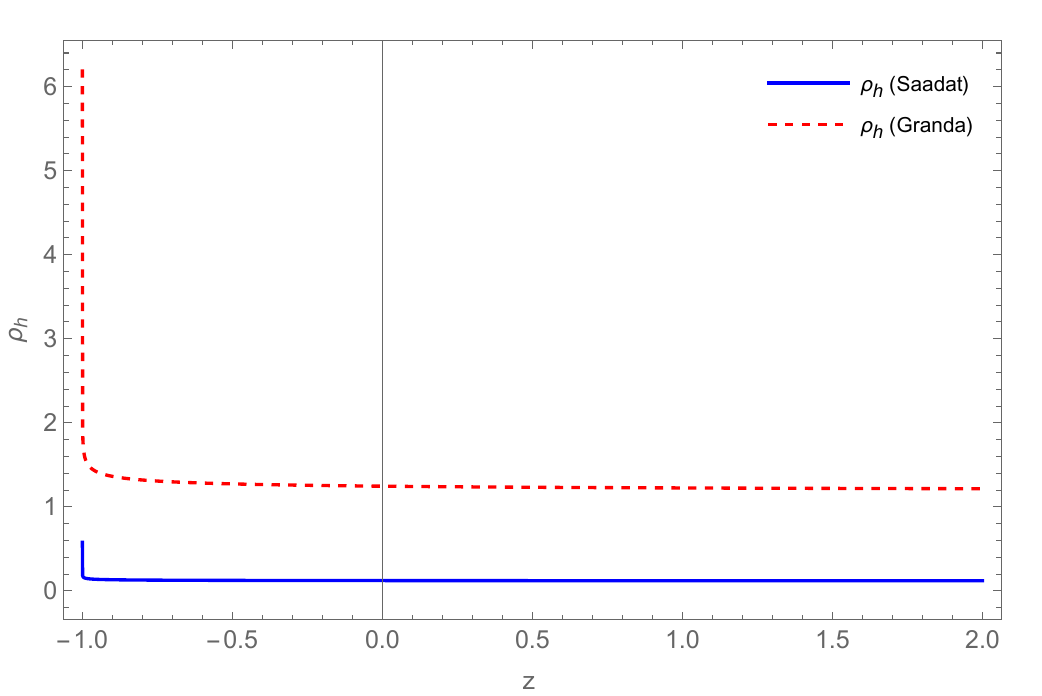}
	\caption{The plot of holographic dark energy density $\rho_h$ \textit{vs.} redshift $z$}
	\label{fig:6}
\end{figure}
Now, the Saadat type energy density of DW $\rho_{d}$ is obtained by using equation \eqref{Eq:32} as
\begin{equation}\label{Eq:34}
	\begin{split}
		\rho_{d}=\left[3(\alpha-d^2)-2\alpha (lt-m)\right]\left(\frac{2}{-lt^2+2mt-2c_{2}}\right)^2 \\ + \left[3-(4n^2-2n)(lt-m)\right]\beta 6^{n-1} \left(\frac{2}{-lt^2+2mt-2c_{2}}\right)^{2 n}
	\end{split}
\end{equation}
Also, from equation \eqref{Eq:33}, the Domain wall energy density $\rho_{d}$ for Granda and Oliveros type is given by
\begin{equation}\label{Eq:35}
	\begin{split}
		\rho_{d}= \left[3 (\alpha -\zeta )-(3 \eta +2 \alpha ) (l t-m)\right] \left(\frac{2}{-lt^2+2mt-2c_{2}}\right)^2 \\ + \left[3-\left(4 n^2-2 n\right) (l t-m)\right] \beta  6^{n-1}  \left(\frac{2}{-lt^2+2mt-2c_{2}}\right)^{2 n}
	\end{split}
\end{equation} 
\begin{figure}[hbt]
	\centering
	\includegraphics[width=0.65\linewidth]{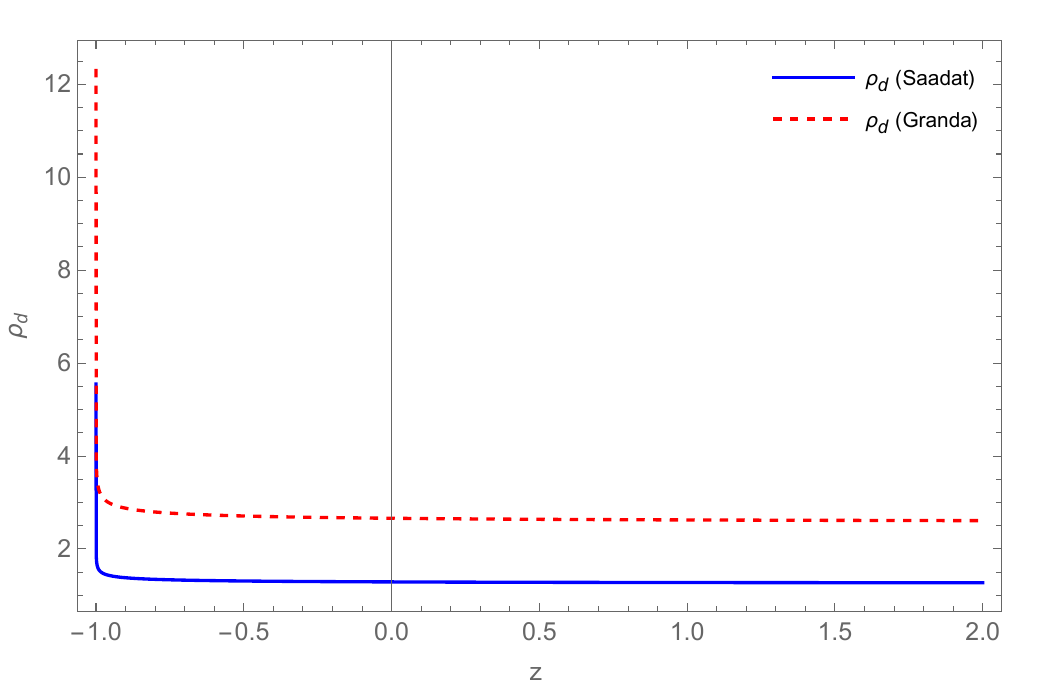}
	\caption{The plot of energy density of domain wall $\rho_d$ \textit{vs.} redshift $z$}
	\label{fig:7}
\end{figure}
From figures \ref{fig:6} and \ref{fig:7}, we observe that the energy densities of HDE and DW are constant and suddenly increasing near $z = -1$. The behaviour for the Saadat \cite{Saadat 2011}, Granda and Oliveros \cite{Granda and Oliveros 2008} types is the same and differs by constant only. The comparison of $\rho_h$ and $\rho_d$ in cases I and II shows that their behaviour is different.
From equation \eqref{Eq:32} and \eqref{Eq:34}, the tension of the domain wall $\sigma_{d}$ regarding Saadat type is obtained as
\begin{equation}\label{Eq:36}
	\begin{split}
		\sigma_{d}=\frac{\left[3\alpha (\gamma -2)-(3d^2+2\alpha (lt-m))(\gamma -1)\right]}{\gamma}\left(\frac{2}{-lt^2+2mt-2c_{2}}\right)^2 \\ + \frac{\left[3(\gamma-2n)-(\gamma -1) \left(4n^2-2n\right)(lt-m)\right]\beta 6^{n-1}}{\gamma} \left(\frac{2}{-lt^2+2mt-2c_{2}}\right)^{2 n}
	\end{split}
\end{equation}
Also, by using equation \eqref{Eq:33} and \eqref{Eq:35}, the domain wall tension $\sigma_{d}$ for Granda and Oliveros type becomes
\begin{equation}\label{Eq:37}
	\begin{split}
		\sigma_{d}=\left[3\alpha \left(1-\frac{2}{\gamma}\right) - \left(1-\frac{1}{\gamma}\right) \left[3\zeta + (2\alpha - 3\eta)(lt - 
		m)\right]\right]\left(\frac{2}{-lt^2+2mt-2c_{2}}\right)^2 \\ + \left[3 - \left(\frac{3n+1}{\gamma}\right)- \left(1-\frac{1}{\gamma}\right)(4n^2 - 2n)(lt-m)\right] \beta 6^{n - 1} \left(\frac{2}{-lt^2+2mt-2c_{2}}\right)^{2 n}
	\end{split}
\end{equation}
\begin{figure}[htb]
	\centering
	\includegraphics[width=0.65\linewidth]{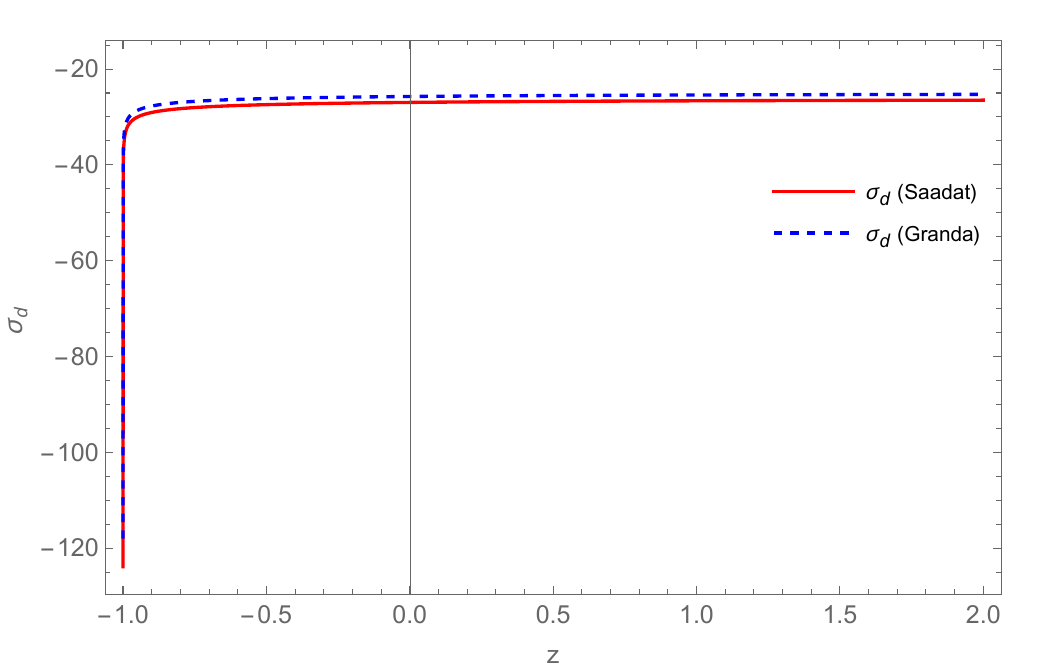}
	\caption{The plot of tension of domain walls $\sigma_d$ \textit{vs.} redshift $z$}
	\label{fig:8}
\end{figure}
From figure \ref{fig:8}, it is found that the tension of the domain wall is negative throughout the evolution of the universe, i.e., the domain wall behaves like invisible matter.

\section{Conclusion}
We have investigated the flat FRW space time using holographic dark energy and domain walls in the framework of the non-metricity $f(Q)$ theory of gravity. For the exact solution, we considered the two different forms of deceleration parameter $q$ in cases I and II. It is observed that the universe is accelerating and expanding, which is confirmed by the cosmological observation data \cite{Riess et al. 1998,Perlmutter et al. 1998,Perlmutter et al. 1999,Spergel et al. 2003} in both cases I and II. The tension of the domain wall is negative, i.e., DW behaves like invisible matter and the universe is dominated by HDE in both cases I and II. According to Zel'Dovich et al. \cite{Zel'Dovich et al. 1974}, DW was present in the early era and vanished at present, which was also observed in our earlier work for FRW spacetime in the $f(R,T)$ theory. Thus, it indicates the effect of HDE or $f(Q)$ theory. In case I, the energy densities $\rho_h$ and $\rho_d$ are decreasing, whereas in case II, the energy densities $\rho_h$ and $\rho_d$ are constant.

\end{document}